\documentclass[useAMS,usenatbib]{mn2e}  
\usepackage{graphicx}
\usepackage{txfonts}
\usepackage{cases}
\title{X-ray and $\gamma$-ray variability of Mrk 421}
\author[Y.G. Zheng et al.]{Y.G. Zheng$^{1}$\thanks{Corresponding author. E-mail: ynzyg@ynu.edu.cn}, S.J. Kang$^{2}$, J. Li$^{1}$\\
       $^{1}$Department of Physics, Yunnan Normal University, Kunming, Yunnan, China\\
       $^{2}$School of Physics, Huazhong University of Science and Technology, Wuhan,China}

\begin{document}
\date{Received date  / Accepted date}

\pagerange{\pageref{firstpage}--\pageref{lastpage}} \pubyear{2014}

\maketitle

\label{firstpage}

\begin{abstract}
We present an ordinary case in the momentum diffusion equation for the electron spectrum evolution and investigate the energy spectra and time dependent properties of flare in Mrk 421 in the frame of a time-dependent one-zone synchrotron self-Compton model. In this model, electrons are accelerated to extra-relativistic energy through the strong magnetic turbulence and evolve with time, and non-thermal photons are produced by both synchrotron and inverse Comtpon scattering off synchrotron photons. Moreover, non-thermal photons during a pre-flare are produced by the relativistic electrons in the steady state and those during a flare are produced by the electrons whose injection rate is changed during some time interval. We apply the model to the energy spectra and time-dependent properties of flares in Mrk 421 by reproducing the pre-burst spectrum of the source and varying the injection rate in the Bohm diffusion (q=1) and the hard sphere approximation (q=2) case, respectively. Our results show that Bohm diffusion case leads to hard photon spectra, and hard sphere case approximation seems to reproduce the energy spectra and the time dependent properties of flare still better.
\end{abstract}


\section{Introduction}
\label{sec:intro}

\label{sec:intro}
Particle acceleration in the relativistic shock is contributed to the first order Fermi acceleration (Fermi-I) at a shock front and the second order Fermi acceleration (Fermi-II) in a downstream region of the shock(Fermi 1949; 1954). Fermi-type acceleration should occur when particles are reflected many times from a distribution of moving magnetic inhomogeneities. In the Fermi-I process, the charged particles crossing the shock front gain the energy. Due to the magnetic inhomogeneities, the particles may be subsequently scattering many times on different sides of the shock front and each time gain the energy. The Fermi-II process in now is identified as an arche-typical form of stochastic acceleration, and the Fermi acceleration term is used in a generic sense to describe stochastic acceleration. During this process, assumes agent of this acceleration is plasma wave turbulence which is expected to be present in non-equilibrium conditions of highly magnetized plasmas. Charged particles, spiraling along magnetic field lines, are then accelerated through resonant interactions with plasma waves. The particles gain or lose energy attributing to the target particles is approaching or receding(Blandford $\&$ Eichler 1987; Gaisser 1990).This problem is often treated in the quasi-linear approximation and leads to Fokker-Planck equation with a diffusion coefficient that magnitude and form depends on the power spectrum and other characteristics of the plasma turbulence(Schlickeiser 1989). {\bf Either Fermi-I mechanism or stochastic acceleration mechanism in a steady state is well known to induce a power law particle energy spectrum with $N(E)\propto E^{-n}$, where $n=1$ is the spectral index. A different spectral index with $n > 1$ can only be achieved if the acceleration process competes with an escape process of particles.} Mostly because second order Fermi acceleration process makes the particles gain or lose energy much longer timescales than the instantaneous first order mechanism(Campeanu $\&$ Schlickeiser 1992; Vainio $\&$ Schlickeiser 1998), it has been neglected for most of the cases. However, in the high velocity turbulent modes, such as $v_{A} \lesssim c$, the efficiency of the second order Fermi acceleration comparable to the first order Fermi acceleration (Virtanen $\&$ Vaino 2005). Furthermore, for the extragalactic jet non-thermal radiation, in which the bulk of radiation is emitted by the particles that have already left the shock discontinuity towards the downstream, we can not neglect its effect on the particle spectrum evolution. Following above issues, Stawarz \& Petrosian (2008) investigate stochastic acceleration of ultrarelativistic electrons by magnetic turbulence. They suggest that the curvature of the high energy segments of the energy spectra, even though they are produced by the same population of electrons, may be substantially different between the synchrotron and inverse Compton components.

Blazars are a special class of active galactic nuclei (AGNs) characterized by a highly variable nonthermal continuum emission extending from radio to $\gamma$-ray bands. The variability on timescales ranging from hours to minutes has been commonly observed at various wavelengths of the electromagnetic spectrum (e.g. B$\rm \ddot{o}$ttcher 2007). In the most extreme cases, the timescales of $\gamma$-rays variability can be as short as a few minutes at very high energy. Such variability has now been observed from several blazars, including Mrk 421 (Gaidos et al. 1996), Mrk 501 (Albert et al. 2007), PKS 2155-304 (Aharonian et al. 2007),1ES 1218-304 (Acciari et al 2009a), 3C 66A (Acciari et al 2009b), 3C 273 (Abdo et al. 2010), PKS 1222+21 (Alsksic et al. 2011), and the prototype object of the class BL Lac (Arlen et al. 2012). The fast flaring pose several challenges to the theoretical model for the $\gamma$-rays emission in blazars. On the one hand, the observed minutes timescale $\rm \gamma$-rays variability implies that the flares are more likely to originate from compact emitting regions that can be most naturally associated with the immediate vicinity of the central engine (Begelman et al. 2008; Giannios et al. 2009; Narayan \& Piran 2012). On the other hand, the $\rm \gamma$-rays emission regions must located at scales $\sim$0.5 pc that $\rm \gamma$-rays can escape attenuation due to external radiation fields (Tavecchio et al. 2011, Nalewajko et al. 2012). Furthermore, for the TeV photons to escape the source, the emitting blob with a Doppler boosting of $\rm \delta\sim$50 towards the observer must be expected. This is much larger that the Lorentz factor $\rm \Gamma\sim$10-20 that is typically inferred for blazar jet from superluminal motions (e.g. Savolainen et al. 2010).

As an open issue, A large number of emission mechanisms responsible for these TeV flares have been put forward to explain the fast flares in individual source, such as, {\bf the relativistic pick-up model (e.g. Gerbig \& Schlickeiser 2007)}, a coherent instability in a compact emission region (e.g. Begelman et al. 2008), reconnection-driven minijets (e.g. Giannios et al. 2009; Giannios 2013), jet deceleration (e.g. Georganoupolos \& Kazanas 2003; Levinson 2007) to wiggles in an anisotropic electron beam directed along the jet (Ghisellini et al. 2009), a magnetized cloud induced stochastic acceleration (Zheng \& Zhang 2011a; Zheng et al. 2011, Zheng et al. 2013), rarefaction waves in magnetized shells (e.g. Lyutikov \& Lister 2010), relativistic turbulence in the jet (e.g. Narayan \& piran 2012), the interaction of the jet with a red giant star (e.g. Barkov et al. 2012), the firehose instability (e.g. Subramanian et al. 2012).

In general, both plasma mechanisms (e.g. Krishan \& Wiita 1994) and beamed radiation (e.g. Crusius-W$\rm \ddot{a}$tzel \& Lesch 1998) are proposed responsible for variability in jets. Steady-state TeV emissions, especially, TeV variability originate from the blazar jet, pose significant challenge. In the lepton model frame, where the electron radiate photons via the synchrotron self-Compton process or the external inverse Compton process, both these mechanisms expect high energy electrons with Lorentz factor $\rm \gamma \sim 10^{4}-10^{5}$. This induced electron acceleration and radiation problems. Motivated by the above issue, in this paper, base on our previous work, we considered an ordinary case in the momentum diffusion equation for the electron spectrum evolution. Our goal is to determin whether an ordinary case in the momentum diffusion equation for the electron spectrum evolution can reproduce the energy spectra and time dependent properties of flare. In \S 2, various physical processes involved in the time dependent particle energy distribution and the nonthermal photon emission are briefly given, including kinetic equations, particle acceleration in turbulent magnetic fields and photon production mechanisms involved in this paper. In \S 3, we apply this model to the variability and the spectral energy distribution (SED) of BL Lacertae object Mrk 421, and conclusions and discussion are given in \S 4. Throughout the paper, we assume the Hubble constant $\rm H_{0}=70km~s^{-1}~Mpc^{-1}$, the matter energy density $\rm \Omega_{M}=0.27$, the radiation energy density $\rm \Omega_{r}=0$ and the dimensionless cosmological constant $\rm \Omega_{\Lambda}=0.73$.

\section{The Model}
\label{sec:model}

We introduce and solve the kinetic equation describing the pile-up mechanism for relativistic electron spectra in a zone within the jet. In the model, As two possible particle acceleration mechanisms are able to produce high energy electrons, Fermi-I and Fermi-II are self-consistent treated in the radiation processes.
\subsection{Kinetic equations}
The description of Fermi type acceleration is in term of isotropic diffusion in momentum space. Schlickeiser (2002) showed that the evolution of the energetic particle distribution can be described by the one dimensional diffusion approximation of the relativistic Vlasov equation:
\begin{equation}
\frac{\partial f(p,t)}{\partial t}=\frac{1}{p^{2}}\frac{\partial}{\partial p}[F(p,f,\frac{\partial}{\partial p}f)]+S(p,t),
\end{equation}
where $f(p,t)$ is the isotropic, homogeneous phase space density with the dimensionless particle momentum $p$, the function $F(p,f,\frac{\partial}{\partial p}f)$ describes the acceleration processes, and $S(p,t)$ describes the catastrophic particle gains and losses. The particle number density $N(p,t)$ is directly related to the phase space density by $N(p,t)=4\pi p^{2}f(p,t)$.

According to particles acceleration via parallel shock front and stochastic acceleration caused by scattering at Alfven waves, the function can be defined (Schlickeiser 1984):
\begin{equation}
F(p,f,\frac{\partial}{\partial p}f)=p^{4}\frac{v_{A}^{2}}{9k_{||}}\frac{\partial f}{\partial p}-p^{3}\frac{v_{s}^{2}}{4k_{||}}f,
\end{equation}
where, $v_{A}$, $v_{s}$ is the Alfven velocity and parallel shock front velocity, respectively. $k_{||}=\frac{1}{3}cL$ is the parallel spatial diffusion coefficient with the light velocity $c$ and parallel spatial mean free path $L$.

Then, we can obtain the equation:
\begin{eqnarray}
\label{Eq.3}
\frac{\partial N(p,t)}{\partial t}&=&\frac{\partial}{\partial p}[-(\frac{v_{s}^{2}}{4k_{||}}+\frac{2v_{A}^{2}}{9k_{||}})p N(p,t)]\nonumber\\&+&\frac{\partial}{\partial p}[\frac{v_{A}^{2}}{9k_{||}}p^{2}\frac{\partial N(p,t)}{\partial p}]+S(p,t),
\end{eqnarray}
The characteristic acceleration timescale $t_{acc}$ is included in Eq. (3):
\begin{equation}
\label{Eq.4}
t_{acc}=(\frac{v_{s}^{2}}{4k_{||}}+\frac{2v_{A}^{2}}{9k_{||}})^{-1},
\end{equation}
Let the parameter $a\approx v_{s}^{2}/v_{A}^{2}$, we can deduce the equation:
\begin{eqnarray}
\label{Eq.5}
\frac{\partial N(p,t)}{\partial t}&=&\frac{\partial}{\partial p}[-\frac{p}{t_{acc}}N(p,t)]\nonumber\\&+&\frac{\partial}{\partial p}[\frac{p^{2}}{(a+2)t_{acc}}\frac{\partial N(p,t)}{\partial p}]+S(p,t),
\end{eqnarray}
As can be seen from Eq.(5), the acceleration process may be quite complex. Such as, the particles may be efficiently accelerated at the shock front, then the energy particles escape into the downstream region of the shock. In this process, the particles are still accelerated by turbulent plasma waves. Therefore, the particles energy may increase enough to make the particle reenter the shock front, and energy spectrum formed by stochastic process may be re-accelerated by the shock (e.g. Katarzynski et al. 2006).

{\bf In a special case, such as in a strong turbulent magnetic fields region, we can neglect the shock acceleration, and then the ratio of shock to stochastic acceleration $a\rightarrow0$. Now that the $t_{acc}$ is not given by Eq. (4) in this extreme case, we should  redefine $t_{acc}^{'}=t_{acc}(a=0)$ in such special case. Making use of the relativistic approximation $p=\gamma$, and let $A(\gamma)=\gamma/t_{acc}^{'}$, $D(\gamma)=\gamma^{2}/(2t_{acc}^{'})$, we rewrite the kinetic equation as:
\begin{equation}
\frac{\partial N(\gamma,t)}{\partial t}=\frac{\partial}{\partial \gamma}\{D(\gamma)\gamma^{2}\frac{\partial}{\partial \gamma}[\frac{N(\gamma,t)}{\gamma^{2}}]\}+S(\gamma, t)
\end{equation}
Eq.(6) is in agreement with the kinetic equation that deduced by diffusion equation approach (e.g. Katarzynski et al. 2006; Tramacere et al. 2011; Zheng \& Zhang 2011a; Zheng et al. 2011; Zheng et al. 2013).}
\subsection{Particle acceleration in turbulent magnetic fields}
Let us further focus on an isotropic Alfvenic turbulence with one dimensional power spectrum $W(k)\propto k^{-q}$ in a finite wave-vector range $k_{min}<k<k_{max}$, where the turbulence magnetic energy density $\int_{k_{min}}^{k_{max}}W(k)dk=\Delta B^{2}/8\pi$, and the turbulence level $\zeta=\Delta B^{2}/B_{0}^{2}$, with $k=2\pi /\lambda$. The $q$ is the power spectrum index, and wave-vector $k_{max}$ and $k_{min}$ correspond to the shortest and longest waves in the system. Using above descried wave spectrum, the momentum diffusion coefficient can be evaluated (Schlickeiser 1989).
\begin{equation}
D(\gamma)\approx\beta_{A}^{2}\zeta \gamma^{2}cr_{g}^{(q-2)}\lambda_{max}^{(1-q)},
\end{equation}
where $\beta_{A}=v_{A}/c$ is the Alfven velocity normalised to the light velocity, and $r_{g}=\gamma m_{e}c^{2}/eB_{0}$ is the gyro-radius of a ultra-relativistic particle. We note that this results is valid for particles with gyro-radii smaller than the correlation length of the turbulence field. The associated parallel spatial mean free path is given by
\begin{equation}
L\approx\frac{1}{3}\zeta^{-1}r_{g}^{(2-q)}\lambda_{max}^{(q-1)},
\end{equation}
This allows us to find the systematic acceleration timescale that contains in Eq.(6) due to stochastic particle-wave interactions,
\begin{equation}
t_{acc}(\gamma)=\frac{\gamma^{2}}{2D(\gamma)}=\frac{\beta_{A}^{-2}}{2}\frac{L}{c},
\end{equation}
The escape timescale due to particle diffusion form the system of turbulent region scale R can be given
\begin{equation}
t_{esc}(\gamma)=\frac{R^2}{k_{||}}=3\zeta R^2c^{-1}\lambda_{max}^{(1-q)}r_{g}^{(q-2)}.
\end{equation}

Since we want to describe the radiative process of the particles, the radiative cooling parameter, $C(\gamma,t)$, that describes the synchrotron and inverse-Compton (IC) cooling of the particles at time t. is introduced.
\begin{equation}
C(\gamma,t)=\frac{4}{3}\frac{\sigma_{T}c}{m_{e}c^{2}}[U_{B}+U_{rad}(\gamma,t)F_{KN}]\gamma^{2},
\end{equation}
where, $\sigma_{T}$ is the Thomson cross section, $m_{e}$ is the electron rest mass, $U_{B}=B^{2}/8\pi$ is the magnetic field energy density, $U_{rad}(\gamma,t)$ is the radiation field energy density, and $F_{KN}$ is Klein-Nishina (KN) effects correction coefficient (e.g. Moderski et al. 2005). {\bf These are in agreement with the SSC energy-loss rate of electron that deduced by Schlickeiser (2009) with the differential KN cross section.} Furthermore, if the diffusion of particles out of the turbulent region is approximated by a catastrophic escape rate $N(\gamma)/t_{esc}$, and there is a source term $Q(\gamma,t)$ description particles injection into the system. Then, one dimensional particle energy distribution is obtained(e.g. Brunetti 2004):
\begin{eqnarray}
\frac{\partial N(\gamma,t)}{\partial t}&=&\frac{\partial}{\partial
\gamma}\{[C(\gamma,t)-A(\gamma,t)]N(\gamma,t) \nonumber\\&+&D(\gamma,t)\frac{\partial
N(\gamma,t)}{\partial \gamma}\}-\frac{N(\gamma,t)}{t_{esc}(\gamma)}+Q(\gamma,t),
\end{eqnarray}

\subsection{Nonthermal photon production}
In the source frame, the total synchrotron emission power per unit volume
\begin{equation}
P_{total}(\nu,t)=\int_{\gamma_{min}}^{\gamma_{max}}P(\nu,\gamma)N(\gamma,t)d\gamma,
\end{equation}
where $P(\nu,\gamma)(erg~s^{-1}~Hz^{-1})$ is the total emitted power per frequency that a relativistic electron in a magnetic field $B$ will radiate fairly broad emission, $N(\gamma,t)$ is the electron number density per unit volume per unit energy at time t. For a distribution of randomly emitter, we can write the synchrotron emission coefficient
\begin{equation}
j(\nu_{syn},t)=\frac{1}{4\pi}P_{total}(\nu,t),
\end{equation}
Thus,
\begin{equation}
j_{syn}(\nu,t)=\frac{\sqrt{3}e^{3}B }{4\pi
m_{e}c^{2}}\int_{\gamma_{min}}^{\gamma_{max}}N(\gamma,t)F(\frac{4\pi
m_{e}c\nu}{3eB_{0}\gamma^{2}})d\gamma,
\end{equation}
where $e$ is the charge of an electron, respectively, $F(x)$ is the modified Bessel functions of $5/3$ order.

Synchrotron emission is accompanied by absorption, in that a photo interacts with an electron, loss its energy. According to a classical scheme of electron-dynamics, we obtain absorption coefficient
\begin{equation}
\alpha(\nu_{syn},t)=-\frac{\sqrt{3}e^{3}B }{8\pi m_{e}^{2}c^{2}}\int_{\gamma_{min}}^{\gamma_{max}}\gamma^{2}F(\frac{4\pi m_{e}c\nu}{3eB_{0}\gamma^{2}})\frac{\partial}{\partial\gamma}[\frac{N(\gamma,t)}{\gamma^{2}}]d\gamma,
\end{equation}
In the spherical geometry structure, the synchrotron intensity is given by many authors(Bloom \& Marscher 1996; Kataoka et al. 1999):
\begin{equation}
I(\nu_{syn},t)=\frac{j(\nu_{syn},t)}{\alpha(\nu_{syn},t)}[1-\frac{2}{\tau^{2}}(1-\tau
e^{-\tau}-e^{-\tau})],
\end{equation}
where $\tau=2R\alpha(\nu_{syn},t)$ is the optical depth of self-absorption.

Follow the approach of Inoue $\&$ Takahara (1996), we calculate self-Compton radiation. We assume a uniform synchrotron intensity in the whole radiation region, corrected for the fact that in reality it decrease along the blob radius(Gould 1979). The effect may be simplified by scaling the intensity of the central point of the blob by a factor 3/4(Kataoka et al. 1999). Thus, the emission coefficient is obtain
\begin{equation}
j(\nu_{ic},t)=\frac{h}{4\pi}\epsilon_{ic}q(\epsilon_{ic},t),
\end{equation}
where $\epsilon_{ic}=h\nu_{ic}/(m_{e}c^{2})$ is the dimensionless particles energy, $q(\epsilon_{ic},t)$ is the differential photon production rate at time t:
\begin{equation}
q(\epsilon_{ic},t)=\int n(\epsilon_{syn},t)d\epsilon_{syn}\int N(\gamma,t)C(\epsilon_{ic},\gamma,\epsilon_{syn})d\gamma,
\end{equation}
where $C(\epsilon_{ic},\gamma,\epsilon_{syn})$ is the Compton kernel given by Jones(1968):
\begin{equation}
C(\epsilon_{ic},\gamma,\epsilon_{syn})=\frac{2\pi r_{e}^{2}c}{\gamma^{2}\epsilon_{syn}}[2\kappa ln\kappa+(1+2\kappa)(1-\kappa)+\frac{(4\epsilon_{syn}\gamma\kappa)^{2}}{2(1+4\epsilon_{syn}\gamma\kappa)}(1-\kappa)],
\end{equation}
where
$\kappa=\epsilon_{ic}/[4\epsilon_{syn}\gamma(\gamma-\epsilon_{ic})]$, and $n(\epsilon_{syn},t)$ is the number density of the synchrotron photons per energy interval, $r_{e}$ is the classical electron radius. $n(\epsilon_{syn},t)$ described by:
\begin{equation}
n(\epsilon_{syn},t)=\frac{3}{4}\frac{4\pi}{hc\epsilon_{syn}}\frac{j(\nu_{syn},t)}{\alpha(\nu_{syn},t)}[1-e^{-\alpha(\nu_{syn},t)R}],
\end{equation}
For a given $\epsilon_{syn}$ and $\gamma$, differential photon production rate $q(\epsilon_{ic})$ can be performed under the range for $\epsilon_{ic}$
\begin{equation}
\epsilon_{syn}\leq \epsilon_{ic}\leq \frac{4\epsilon_{syn}\gamma^{2}}{1+4\epsilon_{syn}\gamma},
\end{equation}
Then, the inverse Compton intensity:
\begin{equation}
I(\nu_{ic},t)=j(\nu_{ic},t)R,
\end{equation}
So, we can obtain the observed flux density at time t, using the Doppler
boosting effect transformation.
\begin{equation}
F_{obs}(\nu,t)=\frac{\pi R^{2}\delta^{3}}{D_{L}^{2}}(1+z)[I(\nu_{syn},t)+I(\nu_{ic},t)],
\end{equation}
where, $\delta$ is the Doppler boosting factor, $z$ is the redshift, and $D_{L}$ is the luminosity distant.
\subsection{$\gamma\gamma$ photon absorption}
At high energies, the Compton photons may produce pairs by
interacting with the synchrotron photons. this process may be
decrease the observed high energy radiation(Coppi $\&$ Blandford
1990; Finke et al. 2009). Katarzynski et al. (2001) analyze the
absorption effect due to pair-production inside the source, they
found that appears almost negligible.

Very high energy($\sim$ TeV) $\gamma$-photons from the source are
attenuated by photons from the extragalactic background light (EBL).
Since we want to apply our result to the Mrk 421, we must take
into account the absorption suffered by TeV photons interacting
with the EBL. The observed photon spectrum, $F_{obs}(\nu)$, of a
source located at redshift $z$ is given by
\begin{equation}
F_{obs}(\nu,t)=e^{-\tau(\nu,z)}F_{in}(\nu,t),
\end{equation}
where $\tau(\nu,z)$ is the absorption optical depth due to
interactions with the EBL(Kneiske et al. 2004; Dwek $\&$ Krennrich
2005).

\section{Application to the variability in Mrk 421}
\label{sec:apply}

The nearby TeV blazar Mrk 421, at a redshift of $z=0.031$, is the first extragalactic source to be detected at energy $\rm E>500$ GeV (Punch et al. 1992), and its $\gamma$-ray flux has been found to be highly variable (Aharonian et al. 1999; Bose et al. 2007; Aleksic et al. 2010). Mrk 421 is reported to undergo one of its brightest flaring episodes on February 10-26, 2010 by SPOL, Swift-XRT, RXTE-PCA, Fermi-LAT, and HAGAR. Especially, a clear variation of flux over a period of seven days is observed in the X-rays and $\gamma$-rays during February 13-19, 2010 (Shukla et al. 2012). This has allowed us to study their spectral, temporal structures, and the correlation of the variability in X-rays and $\gamma$-rays bands. According to the light curve, the spectral of the seven days variability are divided four states as follows: (1) Pre-burst state on February 13-15 2010; (2) X-rays and $\gamma$-rays burst state on February 16 2010; (3) TeV burst state on February 17 2010; and (4) Post burst state on February 18-19 2010. In the following, we show the spectrum and light curve in two extremely case, that is the Bohm diffusion (q=1) and hard sphere approximation (q=2), respectively, and then, we deduce the energy dependent time lag.

\subsection{Photon spectrum and Energy-dependent light curves}
We apply the model described in \S 2 to Mrk 421 and we model its X-ray and $\gamma$-ray variability. In order to do so, we make the following assumptions. (i)The particles are accelerated in a strong turbulent magnetic fields region where the turbulent magnetic fields is $\Delta B\sim B_{0}$  and the turbulent level is $\zeta$=1.  (ii) The relativistic electrons should be in the steady state for $\gamma$-ray emission in the quiescent (pre-burst) state. (iii) Some fresh low energy electrons are injected during the outburst state.
\subsubsection{The case of Bohm diffusion}
In the first test, we assume a higher acceleration efficiency with turbulent power spectrum index q=1. In this case a energy dependent acceleration timescale and escape timescale can be simplified to as follow:
\begin{equation}
t_{acc}(\gamma)=\frac{m_{e}c}{6e}\beta_{A}^{-2}B_{0}^{-1}\gamma,
\end{equation}
\begin{equation}
t_{esc}(\gamma)=\frac{3e}{m_{e}c^{3}}R^{2}B_{0}\gamma^{-1}.
\end{equation}
These are in agreement with the timescales by Tammi \& Duffy (2009). By reason that we are aim at the particles acceleration, in our model, a constant escape timescale, $t_{esc}=R/c=t_{cr}$, is substituted for energy dependent escape timescale.

With above timescales, we can use the time-dependent one-zone SSC solution for spherical geometry. Firstly we search for the steady state solution for electron and photon spectra. Assuming a constant initial electron distribution $N_{\rm ini}(\gamma,0)=8.5\times10^{-15}$ cm$^{-3}$ for $1\leq\gamma\leq2$, we calculate the time evolution of the spectra to the steady state, where the injection rate of the electron population $Q(\gamma)=0.9\times10^{-7}$ cm$^{-3}$ s$^{-1}$ for $1\leq\gamma\leq2$ for all evolution process is assumed. The parameters are used as follows: minimum and maximum Lorentz factors of electrons are  $\gamma_{\rm min}=1$, $\gamma_{\rm max}=10^{8}$, magnetic field strength is $B_{0}=0.0072$ G, Doppler factor is $\delta=13$. The emission region size is constrained by the variability time scales. Variability present in observed data is of the order of one day. We can deduce the co-moving size of the emission region $R\sim c\delta t_{var}/(1+z)=2.25\times10^{15}\delta$ cm. Since the relativistic electrons are in the steady state during the pre-burst of X-rays and $\gamma$-rays. Therefore, we can calculate the pre-burst X-ray and $\gamma$-ray spectrum in the one-zone SSC model using the steady-state electron spectrum. In Figure 1 (a), we show predicted pre-burst spectrum from the X-ray to $\gamma$-ray bands (the solid curve). For comparison, the observed data of Mrk 421 at the X-ray band and the $\gamma$-ray band on February 13-15, 2010 (Shukla et al. 2012) are also shown, where black solid circles with error bars represent the observed values at the pre-burst.

We now consider the properties of the multi-wavelengths flare of Mrk 421 in 2010 February. In order to do it, we use the physical parameters selected above and consider the resulting steady-state spectrum as an initial condition, but we adopt the injection rate function of the electron
population:
\begin{equation}
Q(\gamma, t)=Q_{ini}H(2-\gamma)[27.5\Theta(5.5-t)\Theta(t-5)].
\end{equation}
Here, $\rm Q_{ini}=0.9\times10^{-7}~cm^{-3}~s^{-1}$ is the injection rate constant, $\rm H(\gamma)$ and $\rm \Theta(t)$ are the Heaviside function, and t is evolution time normalized to the light cross time-scale. Under the above assumptions, we, in sequence, reproduce the observed multi-wavelength photon spectrum (the solid curve) of Mrk 421 on February 16, 17, and 18-19, 2010. We show the multi-wavelength photon spectrum in Figure 1 (b), Figure 1 (c), Figure 1 (d), respectively. In order to inspect the evolution of the spectral energy distribution (SED) during the high state, we get together the four state SED in Figure 1 (e). Furthermore we simulated the light curves at energy bands of 0.5-2 KeV, 1.5-12 KeV, 2-20 KeV, 15-50 KeV, 0.2-300 GeV, and $>$ 250 GeV, respectively, and we show the comparisons of the predicted light curves with the multi-wavelength quasi-simultaneous light curve of Mrk 421 (Shukla et al. 2012) in Figure 2. All the model parameters of the one-zone SSC spectra are listed in table 1. It can be seen that, (1) our model can reproduce the multi-wavelength energy spectrum and flare at the X-ray and $\gamma$-ray; (2) During the flare the synchrotron and inverse Compton peak shift towards higher energies; (3) There are steeper photon spectra and stronger IC scattering.

\begin{figure}
\centering
\includegraphics[scale=0.5]{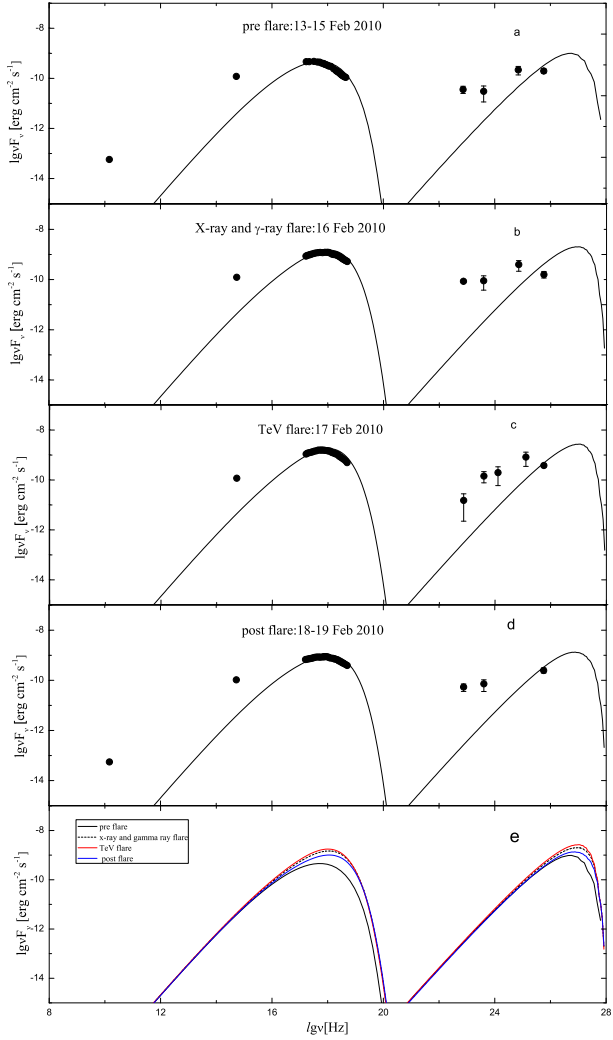}
\caption{Comparisons of predicted multi-wavelength spectra in the case of Bohm diffusion with observed data of Mrk 421 on the February 13-19, 2010. (a) Pre-burst on February 13-15,2010; (b) X-ray and $\gamma$-ray flare on February 16, 2010; (c) TeV flare on February 17, 2010; (d) Post flare on 18-19, 2010; (e) Evolution of the spectral energy distribution during the high state, black solid curve shows the pre-burst state, dash curve shows the X-ray and $\gamma$-ray flare state, red solid curve shows the TeV flare state, and blue solid curve shows the post flare state. Observed data come from Shukla et al. (2012).}
\label{Fig:1}
\end{figure}

\begin{figure}
\centering
\includegraphics[scale=0.5]{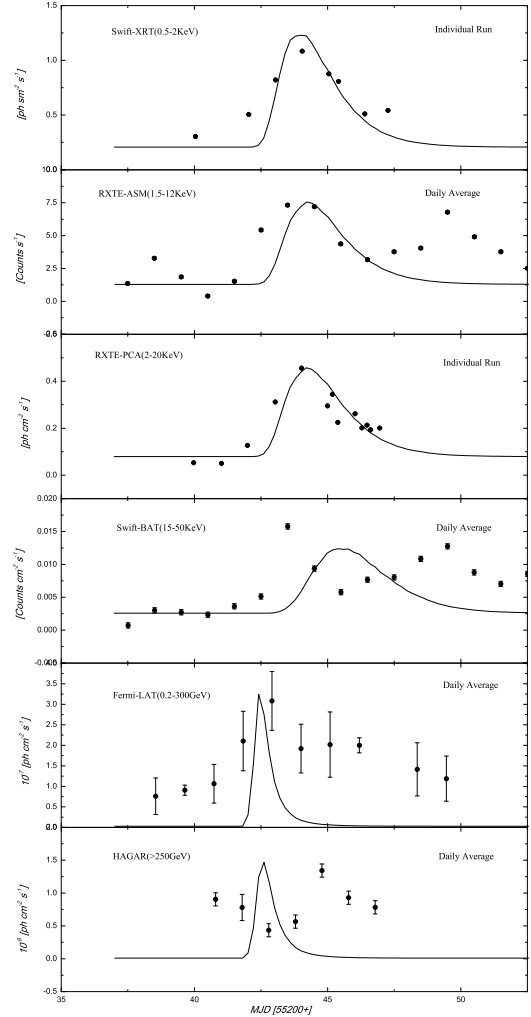}
\caption{Comparisons of simulated light curves in the case of Bohm diffusion (solid lines) with observational light curves (data points) from the Shukla et al. (2012) on the February 13-19, 2010.}
\label{Fig:2}
\end{figure}

\subsubsection{The case of hard sphere approximation}
The second test presented in this section assumes a low acceleration efficiency with turbulence power spectrum index q=2. This scenario leads to a energy independent acceleration timescale, $t_{acc}=\lambda_{max}/(6c\beta_{A}^{2})$, and escape timescale $t_{esc}=3R^{2}/(c\lambda_{max})$. In order to simplify the calculation, we adopt $t_{acc}=t_{esc}=R/c=t_{cr}$.

In this special case, we basically follow the approach of above description to reproduce the photon emission from the Mrk 421 through two states: the steady state spectra and the multi-wavelengths flare state spectra. But we adopt the different parameters and the injection rate function of the electron to the case of Bohm diffusion. For comparison, all the model parameters are also listed in table 1. The injection rate function of the electron population is shown as follow:
\begin{equation}
Q(\gamma, t)=Q_{ini}H(2-\gamma)[20\Theta(5.5-t)\Theta(t-5)].
\end{equation}
To check the validity of the present simulation we again apply our results to the pre-burst state and high energy activity state of Mrk 421. The photon spectra and energy dependent light curves are shown in Figure 3 and Figure 4, respectively. It can be seen that, in the present simulation, the energy spectra and evolution of the peak position are very similar to the above test, but there are a flat photon spectra and weak IC scattering.

\begin{table*}
\begin{minipage}[t][]{\textwidth}
\caption{Physical parameters of the one-zone SSC model spectra}
\label{table1}
\begin{tabular}{llllll}
\hline\hline
 &  \multicolumn{2}{c}{Bohm diffusion} & & \multicolumn{2}{c}{Hard sphere approximation}\\
\cline{2-3} \cline{5-6} \\
 Parameters  & Pre-burst  &  Outburst &  &  Pre-burst  &  Outburst \\
\hline\\
$\gamma_{\rm min}$  & 1.0  & 1.0  &  & 1.0 & 1.0 \\
 $\gamma_{\rm max}$  & $10^{8}$ & $10^{8}$  &   & $10^{7}$  & $10^{7}$ \\
 B~[G]   &  0.0072  &  0.0072  &   & 0.075  & 0.075 \\
 R~[cm] & $2.9\times10^{16}$  & $2.9\times10^{16}$  &   & $1.8\times10^{16}$  & $1.8\times10^{16}$ \\
 $\delta$ & 13 & 13  &   & 8  & 8 \\
 $\rm N_{ini}$~[$\rm cm^{-3}$] &  $8.5\times10^{-15}$  & steady-state spectrum  &   & $1.0\times10^{-4}$  & steady-state spectrum \\
 Q~[$\rm cm^{-3}~s^{-1}$] &  $0.9\times10^{-7}$ & injection rate function  &   &  $1.0\times10^{-4}$ & injection rate function \\
 $\rm \beta_{A}$ &  0.0018  &   0.0018  &  & - & - \\
\hline\\
\end{tabular}
\end{minipage}
\end{table*}

\begin{figure}
\centering
\includegraphics[scale=0.5]{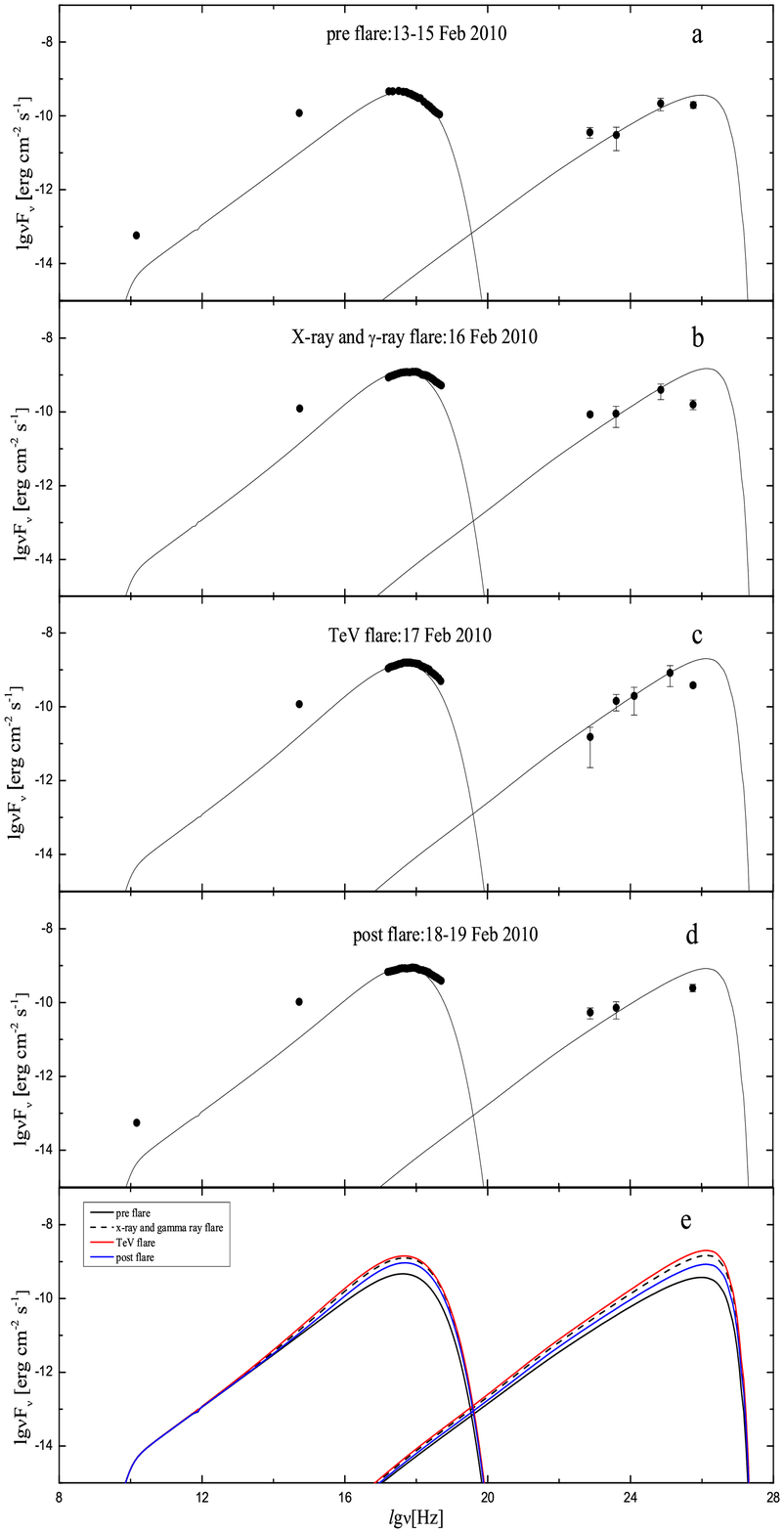}
\caption{Comparisons of predicted multi-wavelength spectra in the case of hard sphere approximation with observed data of Mrk 421 on the February 13-19, 2010. (a) Pre-burst on February 13-15,2010; (b) X-ray and $\gamma$-ray flare on February 16, 2010; (c) TeV flare on February 17, 2010; (d) Post flare on 18-19, 2010; (e) Evolution of the spectral energy distribution during the high state, black solid curve shows the pre-burst state, dash curve shows the X-ray and $\gamma$-ray flare state, red solid curve shows the TeV flare state, and blue solid curve shows the post flare state. Observed data come from Shukla et al. (2012).}
\label{Fig:3}
\end{figure}

\begin{figure}
\centering
\includegraphics[scale=0.5]{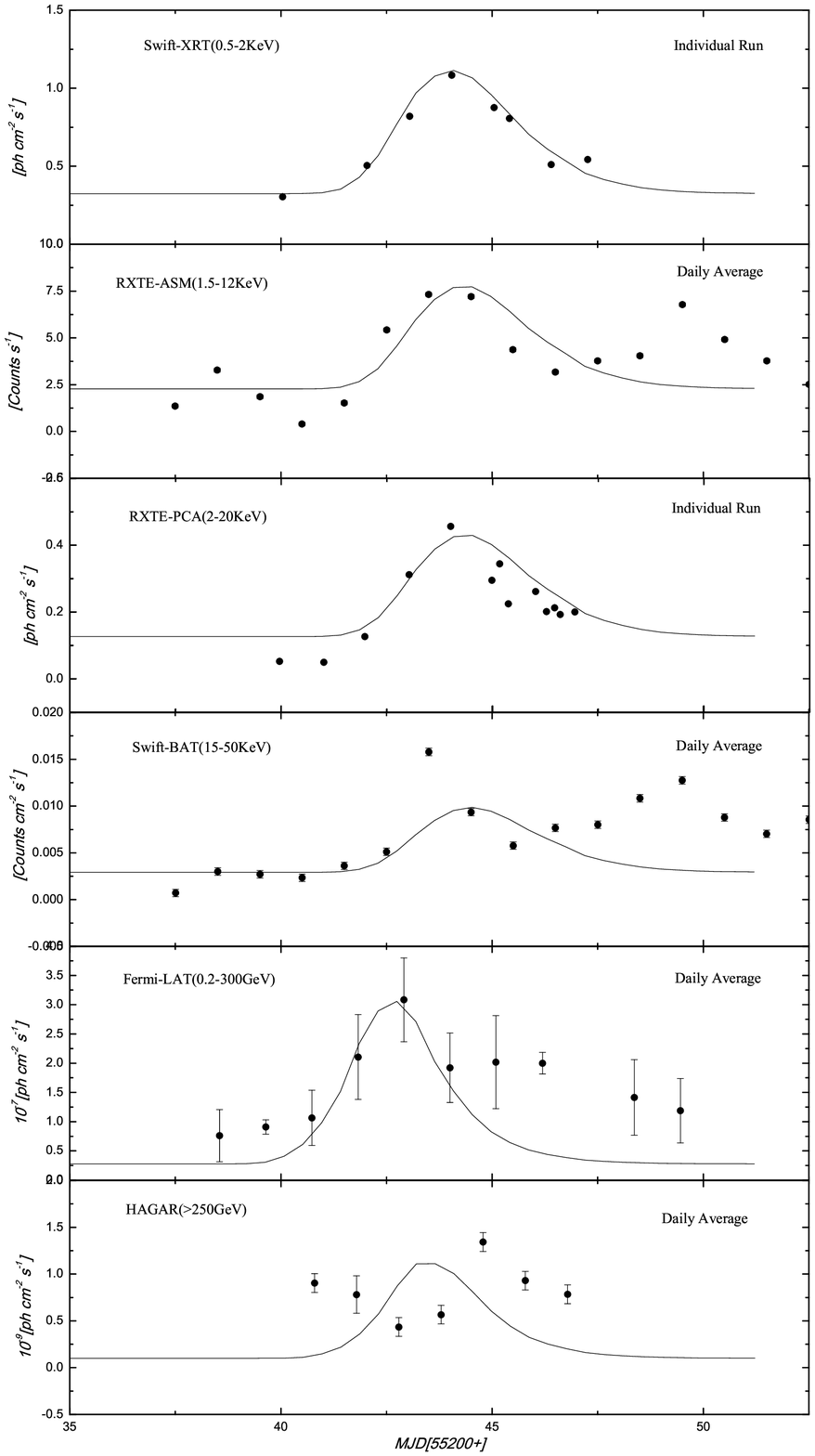}
\caption{Comparisons of simulated light curves in the case of hard sphere approximation (solid lines) with observational light curves (data points) from the Shukla et al. (2012) on the February 13-19, 2010.}
\label{Fig:4}
\end{figure}

\subsection{Time lag}
It is easier to explain soft lags, where lower-energy radiation peaks are later than higher-energy radiation, through high-energy radiating particle cooling and radiating on lower and lower energy. In an opposite sense, hard lags would require some sort of particle acceleration during the flare. Because the possible emission mechanism of X-ray and GeV-TeV $\gamma$-rays are distinct, the time lags are studied on the same regime, such as either synchrotron radiation regime (e.g. Sato et al. 2008) or inverse Compton scattering regime. In this section, using the physical parameters for the energy dependent timescale, we attempt to explain the time lag during giant flare in Mrk421. Noting that the hard sphere approximation leads to a energy independent acceleration timescale, we consider the time lag in the case of Bohm diffusion.

In order to do that, we firstly consider the acceleration timescale, $t_{syn,acc}$, and cooling timescale, $t_{syn,cool}$, in the synchrotron radiation regime. In the source frame, noting that the typical synchrotron radiation frequency of an electron with Lorentz factor $\gamma$, that is averaged over pitch angles is given by $\nu\sim3.7\times10^{6}B_{0}\gamma^{2}$, We can convenient to express $t_{syn,acc}$ and $t_{syn,cool}$ in terms of the observed photon energy $E_{s}$ (in units of KeV) as follows:
\begin{equation}
t_{syn,acc}(E_{s})=2.43\times10^{-3}\beta_{A}^{-2}(1+z)^{3/2}B_{0}^{-3/2}\delta^{-3/2}E_{s}^{1/2}~s,
\end{equation}
\begin{equation}
t_{syn, cool}(E_{s})=3.02\times10^{3}(1+z)^{1/2}B_{0}^{-3/2}\delta^{-1/2}E_{s}^{-1/2}~s.
\end{equation}
Above equations indicated that the cooling timescale $t_{syn,cool}$ in agreement with the timescale by Sato et al. (2008) deduced, through the acceleration timescale $t_{syn,acc}$ different from the timescale by sato et al. (2008) deduced from the shock acceleration with $v_{s}\sim c$(e.g. Inoue \& Takahara 1996).

We now consider the acceleration timescale, $t_{ic,acc}$, and cooling timescale, $t_{ic,cool}$, in the inverse Compton scattering regime. The photon energy $ E_{c}$ (in units of GeV) , that is up scattered synchrotron radiation soft photon with energy $E_{s}$ by an electron with Lorentz factor $\gamma$ is $\sim \gamma^{2}E_{s}$. According to the relations between electron Lorentz factor and synchrotron radiation soft photon with energy $ E_{s}=1.53\times10^{-11}B_{0}\gamma^{2}$, adopting the typical value $U_{rad}=100U_{B_{0}}$, we can obtain $t_{ic,acc}$ and $t_{ic,cool}$ in terms of the observed photon energy $E_{c}$ in the observer's frame as follows:
\begin{equation}
t_{ic,acc}(E_{c})=1.52\times10^{-4}\beta_{A}^{-2}(1+z)^{5/4}B_{0}^{-5/4}\delta^{-5/4}E_{c}^{1/4}~s,
\end{equation}
\begin{equation}
t_{ic, cool}(E_{c})=4.84\times10^{2}(1+z)^{3/4}B_{0}^{-7/4}\delta^{-3/4}E_{c}^{-1/4}~s.
\end{equation}

In order to hunt for the relation between $t_{acc}(E)$ and $t_{cool}(E)$, In Figure 5, we show the photon energy dependent acceleration and cooling timescales, where the synchrotron radiation photon energy and inverse Compton scattering photon energy are normalized to 1 KeV and 1 GeV, respectively. It can be seen from Figure 5 that: (1) there is a relation $t_{acc}(E)<t_{cool}(E)$ for lower energy photons and an opposite sense for higher energy photons; (2) the photons with energy less than the equilibrium energy $E_{eq}$, where there is a relation $t_{acc}(E_{eq})\approx t_{cool}(E_{eq})$, are dominated by the acceleration process.

\begin{figure}
\centering
\includegraphics[scale=0.5]{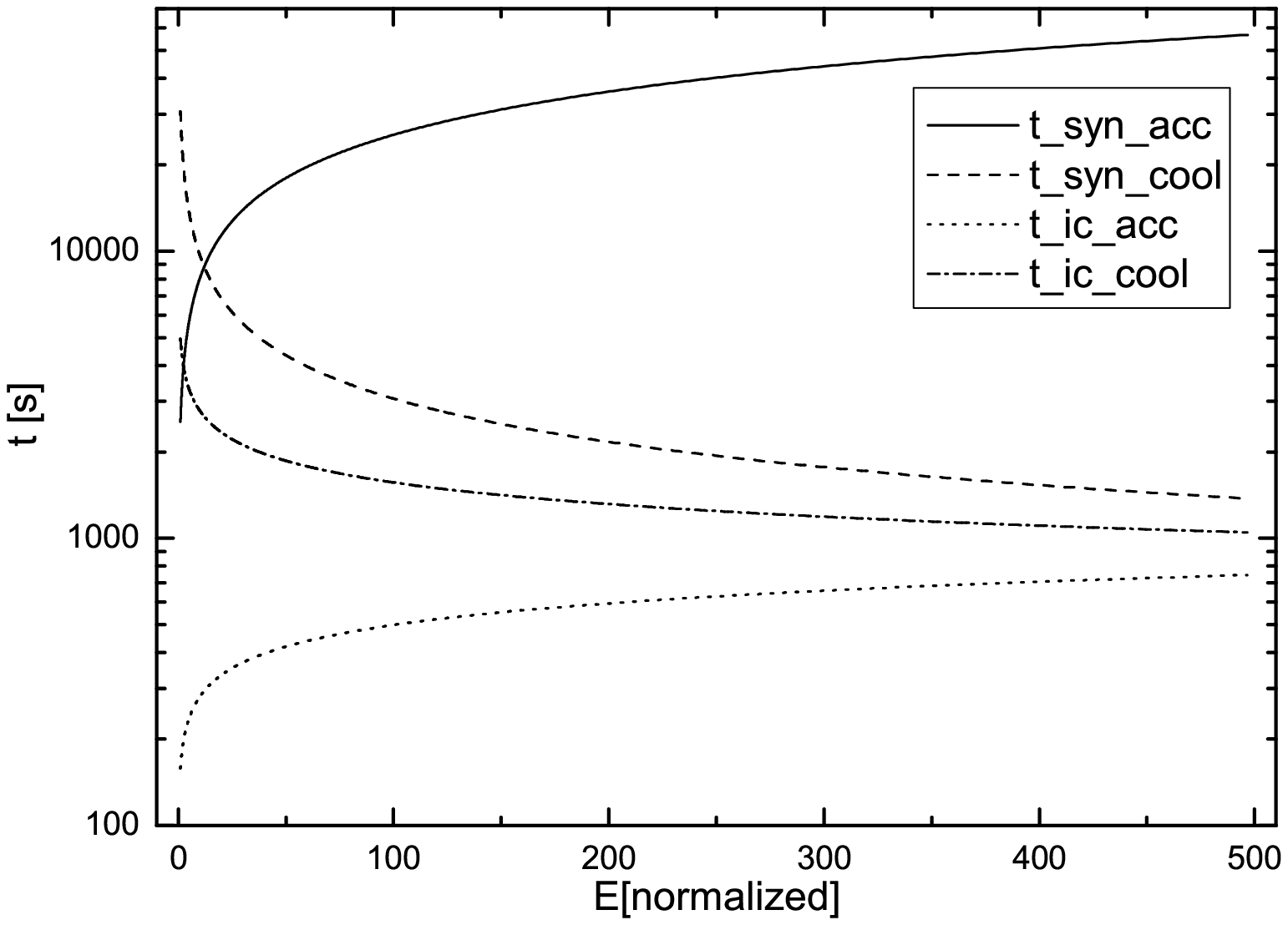}
\caption{The photon energy dependent acceleration and cooling timescales, where the synchrotron radiation photon energy and inverse Compton scattering photon energy are normalized to 1 KeV and 1 GeV, respectively. In generating this plot, we set Alfven velocity $\beta_{A}$=0.001, and we adopt a typical value for the Doppler boosting factor $\delta$=10, and the magnetic field, $B_{0}$=0.1~G.}
\label{Fig:3}
\end{figure}

To ascertain the time of the flare's maximum intensity, we fitted the observed light curves with a function (Norris 1996; Sato et al. 2008):
\begin{equation}
I(t)=\left\{
\begin{array}{ll}
C_{0}+C_{1}\times exp[-(|t-t_{peak}|/\sigma_{r})^{k}]\;\mbox{for $t\le t_{peak}$ }\;,\\
C_{0}+C_{1}\times exp[-(|t-t_{peak}|/\sigma_{d})^{k}] \;\; \mbox{for $t > t_{peak}$ }\;,
\end{array}\right.
\end{equation}
where $C_{0}$ is a offset constant, $t_{peak}$ is the time of the flare's maximum intensity $C_{1}$, $k$ is a measure of pulse sharpness, and $\sigma_{r}$ and $\sigma_{d}$ are the rise and decay time constants. We list the fitting results in table 2. In order to compare the peak time of simulated light curves with the
observed fitting results, in table 2, we also list the model peak time $t_{m, peak}$, that is obtained from the simulated light curves. It can be seen that, (1) the model peak times are in agreement with the observed fitting results; (2) the rise time of the flare  becomes gradually longer with the photon energy increase.

\begin{table*}
\begin{minipage}[t][]{\textwidth}
\caption{X-ray and $\gamma$-ray flare peak time of Mrk 421 in February 2010}
\label{table1}
\begin{tabular}{clccc}
\hline\hline
 &  &\multicolumn{1}{c}{Bohm diffusion} &   & \multicolumn{1}{c}{Hard sphere approximation}\\
\cline{3-3} \cline{5-5}
E & $t_{peak}$ & $t_{m, peak}$ & &$t_{m, peak}$ \\
 &  MJD(55200+)  &  MJD(55200+) & & MJD(55200+) \\
\hline\\
0.5-2 KeV & $43.92\pm0.50^{*}$ & 43.98 & &44.08\\
 1.5-12 KeV  & $44.17\pm0.50$  & 44.23 & &44.15 \\
 2-20 KeV & $44.34\pm0.50$  & 44.26  & &44.30  \\
 15-50 KeV & $44.38\pm0.50$ & 45.37& &44.40\\
 0.2-2 GeV & $43.49\pm0.50$ & - & &- \\
 2-300 GeV &$43.51\pm0.50$ & - & &- \\
 0.2-300 GeV &$42.89\pm0.50$  & 42.42 & &42.75 \\
 $>$250 GeV & $44.69\pm0.50$ & 42.59& &43.64 \\
\hline\\
\end{tabular}
\flushleft{Notes: $^{*}$ We consider the error in 0.50 day level because of the daily average light curve.}
\end{minipage}
\end{table*}

Using the physical parameters in the case of Bohm diffusion for the energy dependent timescale, we can deduce the equilibrium energy $E_{syn, eq}\sim51$ KeV, and $E_{ic, eq}\sim431$ GeV in the synchrotron radiation regime and IC scattering regime, respectively. These indicate that the emission processes in the all of observation bands are dominated by particles acceleration in the emission region. In order to explore the time lag intensively, We interesting to consider a simple scenario in which the rise time of burst is primarily controlled by the acceleration time of the electrons corresponding to observed photon energies (e.g. sato et al. 2008). In this scenario, the hard lag $\tau_{hard}$ simply original from the difference of the $t_{acc}(E)$. So we can write the lag time in the synchrotron regime as
\begin{eqnarray}
\tau_{syn, hard}&=&t_{syn, acc}(E_{s, high})-t_{syn, acc}(E_{s, low})\nonumber\\
&\sim&2.43\times10^{-3}\beta_{A}^{-2}(1+z)^{3/2}B_{0}^{-3/2}\delta^{-3/2}(E_{s, high}^{1/2}-E_{s, low}^{1/2})~s.
\end{eqnarray}
and in the inverse Compton scattering regime as
\begin{eqnarray}
\tau_{ic,hard}&=&t_{ic, acc}(E_{c, high})-t_{ic, acc}(E_{c, low})\nonumber\\
&\sim&1.52\times10^{-4}\beta_{A}^{-2}(1+z)^{5/4}B_{0}^{-5/4}\delta^{-5/4}(E_{c, high}^{1/4}-E_{c, low}^{1/4})~s,
\end{eqnarray}
where $E_{high}$ and $E_{low}$ are the higher and lower photon energies to which the time lag is observed. In Figure 6, we compare expected time lag with the observed time lag. It can be seen from Figure 6, a hard lag would be reproduced while the variability origin of the particles acceleration, through there is large error with the results.

As is easily seen by our results, in the case of Bohm diffusion, we adopt a very low magnetic field, so the IC peak become too high. In this artificial parameter set, the model spectra seem to deviate from the data points apparently, only X-ray spectra seem consistent with the observations. The light curves for various energy bands also seem deviate from the data points especially for 15-50 keV and gamma-ray bands (see Fig.1, Fig.2).  Because that the cooling timescale becomes longer in this parameter set, only the acceleration timescale can be seen by observers. The acceleration timescale causes the specious lag of the hard X-ray. In the case of hard sphere approximation, the model results are better than the Bohm case (see Fig.3, Fig.4). Due to the acceleration timescale does not depend on the electron/photon energy in this special case, we did not discuss the time lag. However, we can clue on the time lag through the model peak time, $t_{m,peak}$, that is listed in table 2. As can be seen that hard lag was shown even in this special case. We argue that the fresh particles with lower energy ($\gamma<2$) are injected into the acceleration region and then are accelerated gradually to the radiation windows by stochastic acceleration. When the acceleration process dominant emission , the hard lag should be expected.

\begin{figure}
\centering
\includegraphics[scale=0.5]{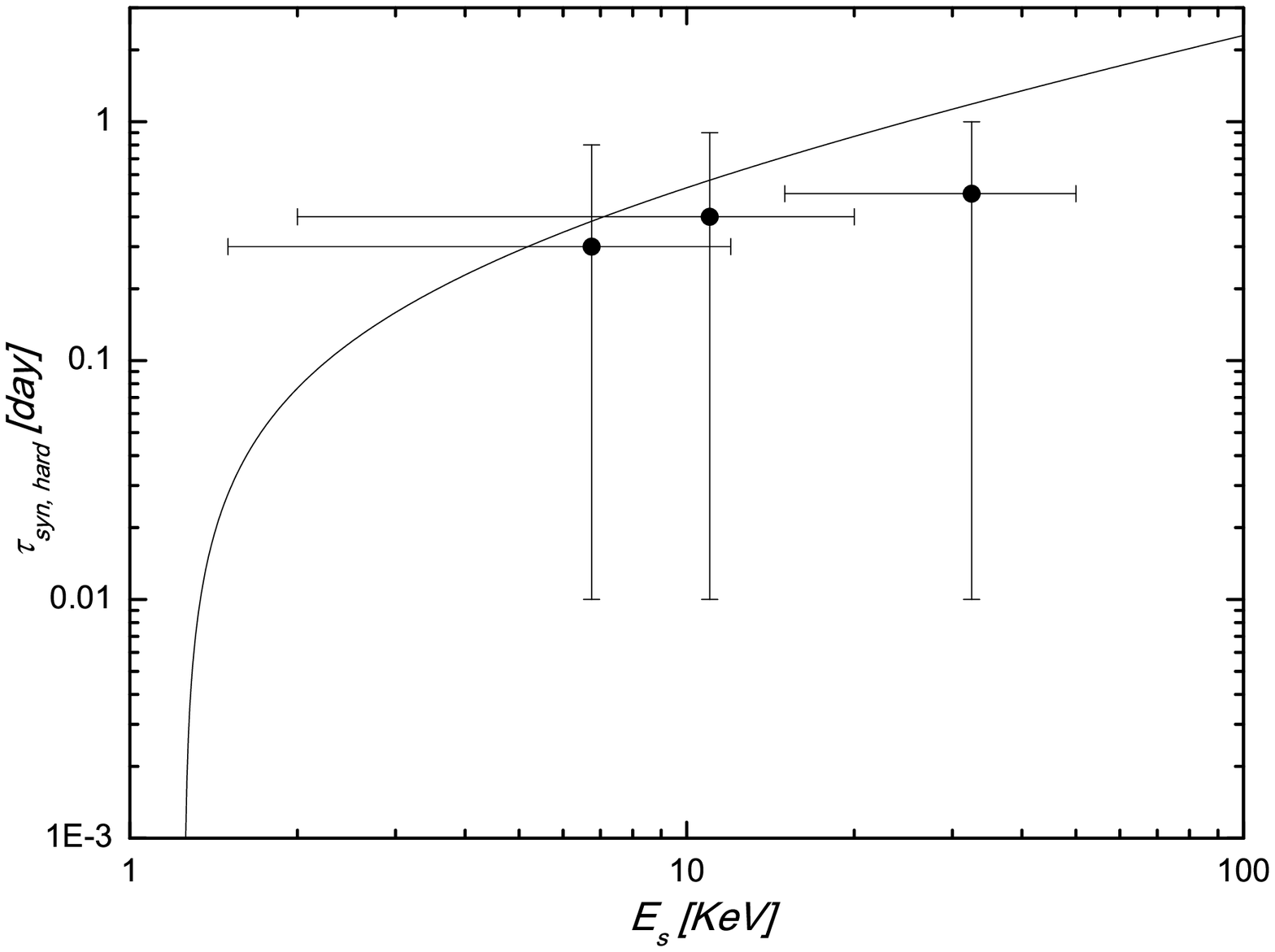}
\includegraphics[scale=0.5]{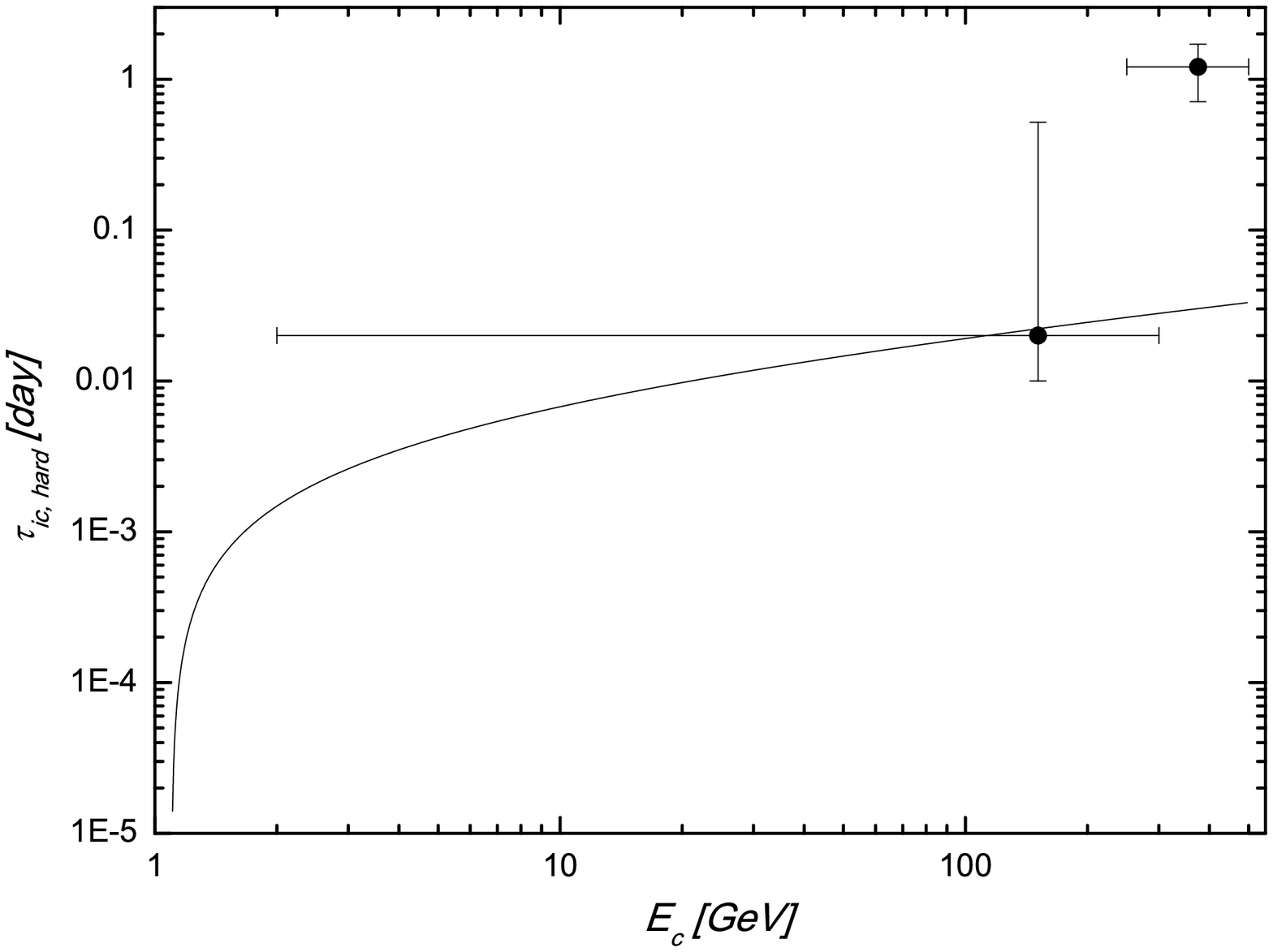}
\caption{Comparison expected time lag with the observed time lag. Top panel shows the case in the synchrotron emission regime with the Lower photon energy $E_{s, low}=1.25$~KeV, and bottom panel shows the case in the inverse Compton scattering regime with the Lower photon energy $E_{c, low}=1.1$~GeV. The data point with the error bar come from the $t_{peak}$ in table 2.}
\label{Fig:3}
\end{figure}

\section{Discussion}
\label{sec:discussion}

In this paper, we try to reproduce the energy spectra and time dependent properties of flare in Blazars, in the context of the time-dependent one-zone SSC model which includes stochastic acceleration of ultra-relativistic particles by strong magnetic turbulence. In this model, particles with low energy are assumed to be injected and then are accelerated to higher energies by stochastic particle-wave interaction, the most important photon targets for IC scattering by relativistic electrons are the synchrotron photons. We study time-dependent properties of flares by reproducing the pre-burst spectrum of the source and varying the injection rate in the Bohm diffusion (q=1) and the hard sphere approximation (q=2) case, respectively. Our results show that Bohm diffusion case leads to hard photon spectra, and hard sphere case approximation seems to reproduce the energy spectra and the time dependent properties of flare still better.

The calculations assumed a constant escape timescale, $t_{esc}=R/c$, to substitute for the energy dependent escape timescale. It can be seen that, in the range of whole Lorentz factor, the energy dependent escape timescale is more than the constant escape timescale. That is, the constant escape timescale leads to higher escape efficiency and softer spectra. Now that we account for the stochastic acceleration, the spectra became too hard (e.g. see Lefa, Aharonian\& Rieger 2011; Asano et al. 2014), especially, we obtain stronger IC scatter in the Bohm case. Even though, adopting same weak magnetic field, Abdo et al. (2011) deduced a lower IC peak with the steady spectrum. It is considered that higher acceleration efficiency in the Bohm case make the more low energy particles to pile-up near the radiation windows, the stronger IC scattering can be expected. With the acceleration efficiency decreasing (hard sphere case), the observed spectra can be fitted by the soften model spectrum.

There are two scenarios for explaining the intrinsic variability. One assumes that the observed variations originate from the geometry of emitting sources (e.g., Camenzind \& Krockenberger 1992; Gopal-Krishna \& Wiita 1992). The other assumes that the variability is generated by the change of the emission condition. A typical example is that fresh particles are injected into the acceleration region and then are accelerated (e.g., Blandford \& Konigl 1979; Marscher \& Gear 1985; Celotti et al. 1991; Kirk et al. 1998). Chen et al. (2011) investigate a set of simple scenarios where the variability is produced by injection of relativistic electrons as a shock front crosses the emission region, and suggest that different phases of activity may occur in the same region. In order to reproduce the multi-wavelength spectra and variability, the model changed only the injection rate for outburst state. This implies that the outbursts are triggered by a magnetized cloud in same region, in which there is a magnetic field and bulk factor in consonance with the pre-burst state. It should be noted that when the magnetized cloud comes into being in the jet in which the local plasma density is enhanced, the number of particles increase as an avalanche occurs in the jet, and the injection rate can be expected to change. In this scenario, abundance of energy particles are accelerated to higher energy, the hard photon spectra should be expected.

As can be seen that the radio and optical portion of the SED is not well fit by this single zone model. Although the multi-wavelength observed results indicate that the radio and optical variability of Mrk 421 are accompanied by strong X-rays and $\gamma$-rays variability (Shukla et al. 2012). It is likely that a different population of particles is responsible for the synchrotron emission at these energies (e.g. Blazejowski et al. 2005; Horan et al. 2009). Since the observed energy bands by Fermi-LAT and HAGAR far from the equilibrium energy $E_{eq}$, the high acceleration efficiency makes the electron energy reach to the radiation window rapidly. So, in the Bohm diffusion case, we can expect a quasi-monochromatic light curves in the Fermi-LAT and HAGAR band, respectively.

It should be noted that if the variability timescale is faster than the cooling timescale, the radiation from accelerated particles would show a hard lag (Albert et al. 2008). In this view, Mastichiadis \& Moraitis (2008) argue that the observed hard lag features can be explained with a physically modification, that is, allowing the particles to accelerate gradually. In the Bohm diffusion scenario, assuming low energy electron injection and stochastic acceleration by the turbulent magnetic field, our calculations predicted a energy dependent hard lag flaring activity. Actually, the hard lag was seen even in the hard sphere approximation case (Zheng \& Zhang 2011a). Similar result was also predicted by simulations of stochastic acceleration in relativistic shocks (e.g. Virtanen \& Vainio 2005). As a result, hard lag would be more expected in lower energy band, now that the photons energy is less than the equilibrium energy. However, the observational results show an opposite sense. It has been suggested that a hard lag is observed at energies closer to the equilibrium energy (Kirk et al. 1998). The fact that a hard lag is actually observed in some blazars, such as Mrk 421 (e.g. Zhang 2002; Ravasio et al. 2004), Mrk 501 (Albert et al. 2007) and 1ES 1218+304 (Sato et al. 2008).  We argue that the observational time bins are longer than the acceleration time-scales probably means that, in practice, there is a loss of radiative lifetime.

It is clear that the power spectrum index $q (1\leq q\leq 2)$ plays an important part in determining a electron spectra shape and evolution (e.g. Zheng \& Zhang 2011b). In our results, the behaviours of the multi-wavelength spectra between in the Bohm diffusion and in the hard sphere approximation case are a little different. The energy spectra shows steeper shape and stronger IC scattering in the Bohm diffusion case and the opposite behaviours in the hard sphere approximation case. It is considered that the higher particles acceleration efficiency leads to steeper photon spectra and stronger IC scattering in the Bohm diffusion case than in the hard sphere approximation case. As an open issue, other power spectra case, such as the Kreichnan turbulence ($q=3/2$) and Kolmogorov turbulence ($q=5/3$) should be investigated.

In the model, we must note that as the acceleration particles need electromagnetic turbulence for scattering and energy gain (e.g. see, Tammi \& Duffy 2009). It is obvious that suitable turbulence either has to exist in the pre-shock plasma or be generated at the shock by the particles themselves (Bell 2004). The numerical results show that sufficient turbulence for fast acceleration could be provided by nonlinear turbulence (Reville et al. 2006). Generally, stochastic acceleration occurs wherever there are turbulent magnetic fields and can spread to an extended region; the size is determined by the turbulence generation and decay rates (Campeanu \& Schlickeiser 1992). Some recent observations suggests that the stochastic acceleration is seen in the observations (Katarzynski et al. 2006; Tramacere et al. 2007; B$\ddot{o}$ttcher et al. 2008; Tramacere et al. 2011). The model presented here contains the stochastic acceleration process. In order to reproduce the multi-wavelength energy spectra and variability, we assume shorter acceleration and escape timescales in the hard sphere approximation case than that are adopted by other investigators (see, e.g., Kirk et al. 1998; Mastichiadis \& Moraitis 2008). Above assumptions can lead to stronger turbulence with Alfven velocity $\rm \beta_{A}\approx 0.7$, and longer waves with $\rm \lambda_{max}=3R$.  These imply that there is a higher acceleration rate and a lower escape rate, and make more particles accelerate to high energy rapidly. Assuming suitable parameters, we obtain excellent fits to the observed spectra, variability and energy dependent time lag of BL Lac object Mrk 421. This indicates that the variable characteristic can be explained by stochastic acceleration energy particles. Given the complexity of the flaring activity in blazars, this requires more detailed observations and that the issue remains open.

\section*{Acknowledgments}
We thank the anonymous referee for valuable comments and suggestions.
This work is partially supported by the National Natural Science Foundation of China under grants 11178019, U1231203, Science and Technology in support of Yunnan Province Talent under grants 2012HB014, and the Natural Science Foundation of Yunnan Province under grants 2011FB041. This work is also supported by the Science Foundation of Yunnan educational department (grant 2012Z016).


\end{document}